\documentclass[aps,pra,reprint,preprintnumbers,amsmath,amssymb,superscriptaddress,showpacs,showkeys,longbibliography]{revtex4-1}
\usepackage[latin1]{inputenc}  
\usepackage[english]{babel}
\usepackage{amsmath}	
\usepackage{amsfonts}
\usepackage{amssymb}
\usepackage{graphicx}	
\usepackage{siunitx}	
\usepackage{subfigure}

\usepackage{placeins}

\usepackage[pdftex]{hyperref} 

\usepackage[draft]{changes}

\hypersetup{%
  colorlinks=true,   
  citecolor=blue,
  urlcolor=blue,
  linkcolor=black,
  pdfpagemode=UseNone,  
  pdfstartview=FitH} 

\hypersetup{
  pdftitle={Competing signal paths and nonlinearity engineering in coupled superconducting resonators},
  pdfauthor={Fischer, Michael},
  pdfsubject={a},
  pdfkeywords={a}
}

\begin{document}

\title[]{In-situ tunable nonlinearity and competing signal paths in coupled superconducting resonators}

\author{Michael Fischer}
\email[]{michael.fischer@wmi.badw.de}
\affiliation{Walther-Mei{\ss}ner-Institut, Bayerische Akademie der Wissenschaften, 85748 Garching, Germany }
\affiliation{Physik-Department, Technische Universit\"{a}t M\"{u}nchen, 85748 Garching, Germany}
\author{Qi-Ming Chen}
\affiliation{Walther-Mei{\ss}ner-Institut, Bayerische Akademie der Wissenschaften, 85748 Garching, Germany }
\affiliation{Physik-Department, Technische Universit\"{a}t M\"{u}nchen, 85748 Garching, Germany}
\affiliation{Munich Center for Quantum Science and Technology (MSQCT), Schellingstra{\ss}e 4, 80799 M\"{u}nchen, Germany}
\author{Christian Besson}
\affiliation{Walther-Mei{\ss}ner-Institut, Bayerische Akademie der Wissenschaften, 85748 Garching, Germany }
\affiliation{Physik-Department, Technische Universit\"{a}t M\"{u}nchen, 85748 Garching, Germany}
\author{Peter Eder}
\affiliation{Walther-Mei{\ss}ner-Institut, Bayerische Akademie der Wissenschaften, 85748 Garching, Germany }
\affiliation{Physik-Department, Technische Universit\"{a}t M\"{u}nchen, 85748 Garching, Germany}
\author{Jan Goetz}
\affiliation{Walther-Mei{\ss}ner-Institut, Bayerische Akademie der Wissenschaften, 85748 Garching, Germany }
\affiliation{Physik-Department, Technische Universit\"{a}t M\"{u}nchen, 85748 Garching, Germany}
\author{Stefan Pogorzalek}
\affiliation{Walther-Mei{\ss}ner-Institut, Bayerische Akademie der Wissenschaften, 85748 Garching, Germany }
\affiliation{Physik-Department, Technische Universit\"{a}t M\"{u}nchen, 85748 Garching, Germany}
\author{Michael Renger}
\affiliation{Walther-Mei{\ss}ner-Institut, Bayerische Akademie der Wissenschaften, 85748 Garching, Germany }
\affiliation{Physik-Department, Technische Universit\"{a}t M\"{u}nchen, 85748 Garching, Germany}
\affiliation{Munich Center for Quantum Science and Technology (MSQCT), Schellingstra{\ss}e 4, 80799 M\"{u}nchen, Germany}
\author{Edwar Xie}
\affiliation{Walther-Mei{\ss}ner-Institut, Bayerische Akademie der Wissenschaften, 85748 Garching, Germany }
\affiliation{Physik-Department, Technische Universit\"{a}t M\"{u}nchen, 85748 Garching, Germany}
\affiliation{Munich Center for Quantum Science and Technology (MSQCT), Schellingstra{\ss}e 4, 80799 M\"{u}nchen, Germany}
\author{Michael J. Hartmann}
\affiliation{Friedrich-Alexander University Erlangen-N\"{u}rnberg (FAU), Department of Physics, 91058 Erlangen, Germany}
\affiliation{Max Planck Institute for the Science of Light, 91058 Erlangen, Germany}
\author{Kirill G. Fedorov}
\affiliation{Walther-Mei{\ss}ner-Institut, Bayerische Akademie der Wissenschaften, 85748 Garching, Germany }
\affiliation{Physik-Department, Technische Universit\"{a}t M\"{u}nchen, 85748 Garching, Germany}
\author{Achim Marx}
\affiliation{Walther-Mei{\ss}ner-Institut, Bayerische Akademie der Wissenschaften, 85748 Garching, Germany }
\author{Frank Deppe}
\affiliation{Walther-Mei{\ss}ner-Institut, Bayerische Akademie der Wissenschaften, 85748 Garching, Germany }
\affiliation{Physik-Department, Technische Universit\"{a}t M\"{u}nchen, 85748 Garching, Germany}
\affiliation{Munich Center for Quantum Science and Technology (MSQCT), Schellingstra{\ss}e 4, 80799 M\"{u}nchen, Germany}
\author{Rudolf Gross}
\email[]{rudolf.gross@wmi.badw.de}
\affiliation{Walther-Mei{\ss}ner-Institut, Bayerische Akademie der Wissenschaften, 85748 Garching, Germany }
\affiliation{Physik-Department, Technische Universit\"{a}t M\"{u}nchen, 85748 Garching, Germany}
\affiliation{Munich Center for Quantum Science and Technology (MSQCT), Schellingstra{\ss}e 4, 80799 M\"{u}nchen, Germany}

\date{\today}

\begin{abstract}
We have fabricated and studied a system of two tunable and coupled nonlinear superconducting resonators. The nonlinearity is introduced by galvanically coupled dc SQUIDs. We simulate the system response by means of a circuit model, which includes an additional signal path introduced by the electromagnetic environment. Furthermore, we present two methods allowing us to experimentally determine the nonlinearity. First, we fit the measured frequency and flux dependence of the transmission data to simulations based on the equivalent circuit model. Second, we fit the power dependence of the transmission data to a model that is predicted by the nonlinear equation of motion describing the system. Our results show that we are able to tune the nonlinearity of the resonators by almost two orders of magnitude via an external coil and two on-chip antennas. The studied system represents the basic building block for larger systems, allowing for quantum simulations of bosonic many-body systems with a larger number of lattice sites.
\end{abstract}

\maketitle

\section{Introduction}
The field of analog quantum simulation~\cite{Feynman2018} has opened up the possibility to experimentally simulate quantum phenomena without the need for universal quantum computing~\cite{Britton2012,Lloyd1996}. Especially in the field of quantum many-body physics, where calculations with classical computational approaches are inefficient, analog quantum simulations may lead to a better understanding of the underlying models~\cite{Noh2017,Bloch2008,Ciuti2013}. One of the very promising platforms for such simulations is circuit quantum electrodynamics (QED), where the quantum behaviour of superconducting circuits is used to emulate the quantum system under test~\cite{Hartmann2016,Houck2012,Carusotto2020,Fitzpatrick2016,Lang2011,Collodo2019}. Circuit QED allows for a large degree of design flexibility~\cite{Deppe2008,Niemczyk2010,Baust2015}, experimental control~\cite{Baust2015,Goetz2017,Wulschner2016}, \textit{in-situ} tunability of essential parameters~\cite{Goetz2018,Schwarz2013}, and scalability~\cite{Houck2012}. 

A particular example of a quantum many-body system showing rich physical effects to be investigated with quantum simulations is the Bose-Hubbard model (BHM)~\cite{Riccardo2019}. It describes the behavior of interacting bosons on a lattice. It has been shown that this model can be simulated in a bottom-up approach by a network of coupled superconducting resonators, each equipped with a direct current superconducting interference device (dc SQUID) at its current antinode~\cite{Bourassa2012,Leib2012,Leib2010, Leib20132,Eichler2014}. The coupling between SQUID and resonator creates polaritons, which are quasi-particles formed by a superposition of the photonic excitation of the resonator and the matter-like excitation of the SQUID. These polaritons can then be used for simulating the bosonic interaction of the BHM. The implementation of the BHM with superconducting circuits represents an open quantum system, where the particle number at each site can be controlled by the interplay between externally applied microwave drives and local dissipation channels. The natural access to this driven-dissipative regime of the BHM distinguishes superconducting-circuit implementations from cold-atom implementations~\cite{Bloch2008,Hartmann2016} and is expected to exhibit exciting novel phases of light. A prototypical example in this context is the theoretical prediction of polariton crystallization~\cite{Hartmann2010}. Experimentally, a photon ordering phase transition~\cite{Collodo2019} and a dissipation-driven transition between localization and delocalization~\cite{Raftery2014} have been shown. Furthermore, studies on phase transitions in large scale circuits without individual control of each lattice site~\cite{Fitzpatrick2016} exist and a single phase, i.e. the Mott insulator phase of photons, could be stabilized in a lossy system~\cite{Ma2019}. A key prerequisite for the application of artificial circuits in quantum simulation experiments is the ability to accurately design, determine and control the relevant circuit parameters. Hence, a detailed understanding and modeling of quantum circuits consisting of coupled nonlinear superconducting resonator as well as their interaction with the environment~\cite{Eder2018} is of large importance. 

In this work, we address this topic by investigating a system of two coupled resonators with a weak, but tunable nonlinearity. So far, this regime has mostly been investigated in the quite different context of parametrically driven circuits~\cite{Zhong2013,Pogorzalek2017,Fedorov2018}. We discuss how to set up a circuit model for the characterization of such a system in the presence of a spurious environment whose microscopic origin does not need to be exactly known. We show that this environment can be modeled with a spurious parallel signal path giving rise to Fano-like resonances. In this way, we gain access to the full parameter space of the coupled systems in a controlled way. Consequently, we can investigate a key property of our system: the nonlinearity of the resonators. Specifically, we employ two different characterization techniques including a direct measurement. We show that the nonlinearity of our resonators can be tuned \textit{in situ} from values much smaller to values larger than the resonator-resonator coupling. In this way, we provide a technique for a controlled access to promising parameter regimes for future quantum simulations.
\section{Sample \& Experimental Setup}

\begin{figure}
\centering
\includegraphics[width = 0.45\textwidth]{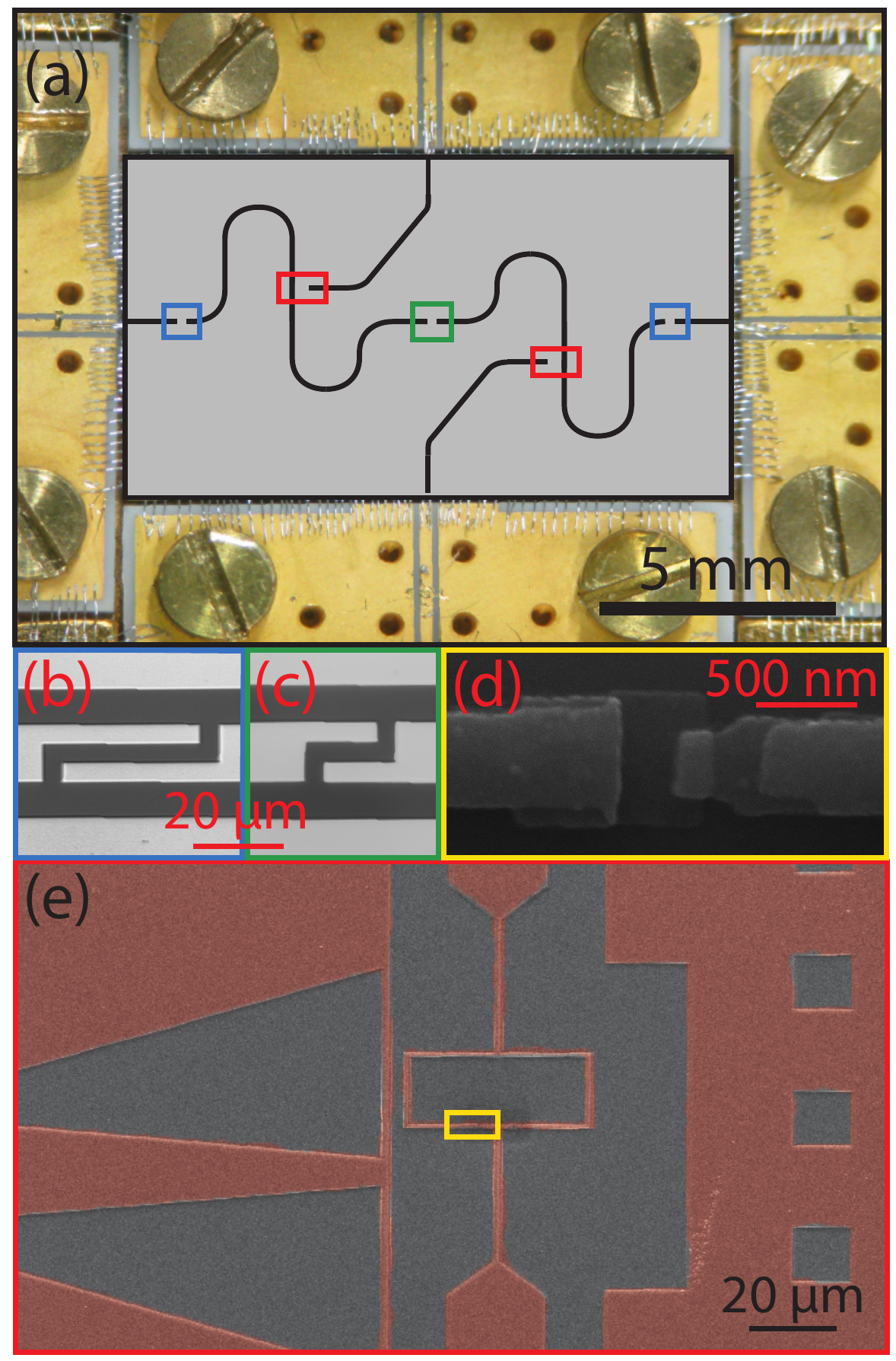}
\caption{(a) Sketch of the two-resonator sample chip (black lines: resonators, feed lines and antennas) mounted into a photograph of the sample holder. The ground planes are connected across all on-chip CPW structures by Al wire bonds (not shown in the sketch for clarity) spaced in regular intervals. Colored rectangles indicate zoom-in views.  (b) \SI{40}{\micro\metre} long finger capacitor to couple the \SI{7420}{\micro\metre} long resonator to the external feedline. The width of the inner conductor of the waveguide is \SI{13.2}{\micro\metre} and gaps between inner and outer conductor are \SI{8}{\micro\metre} each. (c) \SI{20}{\micro\metre} finger capacitor connecting the two coplanar waveguide resonators. (d) Zoom-in view showing one of the Josephson junctions of the (e) dc SQUIDs (false color micrograph). The SQUIDs are galvanically coupled to the inner conductor of each resonator at the current antinode. The structure on the left of the SQUID loop is one of the on-chip antennas.} 
\label{fig:samplepic}
\end{figure}

In our experiments, we use a sample consisting of two weakly-coupled superconducting resonators fabricated on a \SI{525}{\micro\metre} thick silicon chip using aluminum technology. The whole metal layer including the Josephson junctions is fabricated using double-angle shadow evaporation and lift-off. The overall thickness of the aluminum layer is \SI{140}{\nano\metre}. A photo and micrographs of the sample are shown in Fig.~\ref{fig:samplepic}. The main part of the superconducting circuit is formed by a series connection of two capacitively coupled coplanar waveguide resonators, which are each intersected by a dc SQUID. The area of the SQUID loop is $A_\text{SQUID} = \SI[product-units = brackets-power]{10.5 x 24.5}{\micro\metre}$. By design, the two Josephson junctions differ in size in order to flatten the flux dependence of the SQUID critical current. In this way, the sensitivity of the system to  magnetic flux is decreased. For our junctions, we measure an asymmetry parameter  $d = \left(I_\text{c1}-I_\text{c2}\right)/\left(I_\text{c1}+I_\text{c2}\right) \simeq 0.13$  for both SQUIDs (see Sec.~\ref{sec:circuitmodel}). Here, $I_\text{c1}$ and $I_\text{c2}$ are the critical currents of the two Josephson junctions in the dc SQUID. We can tune the two resonators in a frequency range between \SI{5.67}{\giga\hertz} and \SI{7.14}{\giga\hertz} by an applied magnetic flux.
\begin{figure}
\centering
\includegraphics[width = 0.4\textwidth]{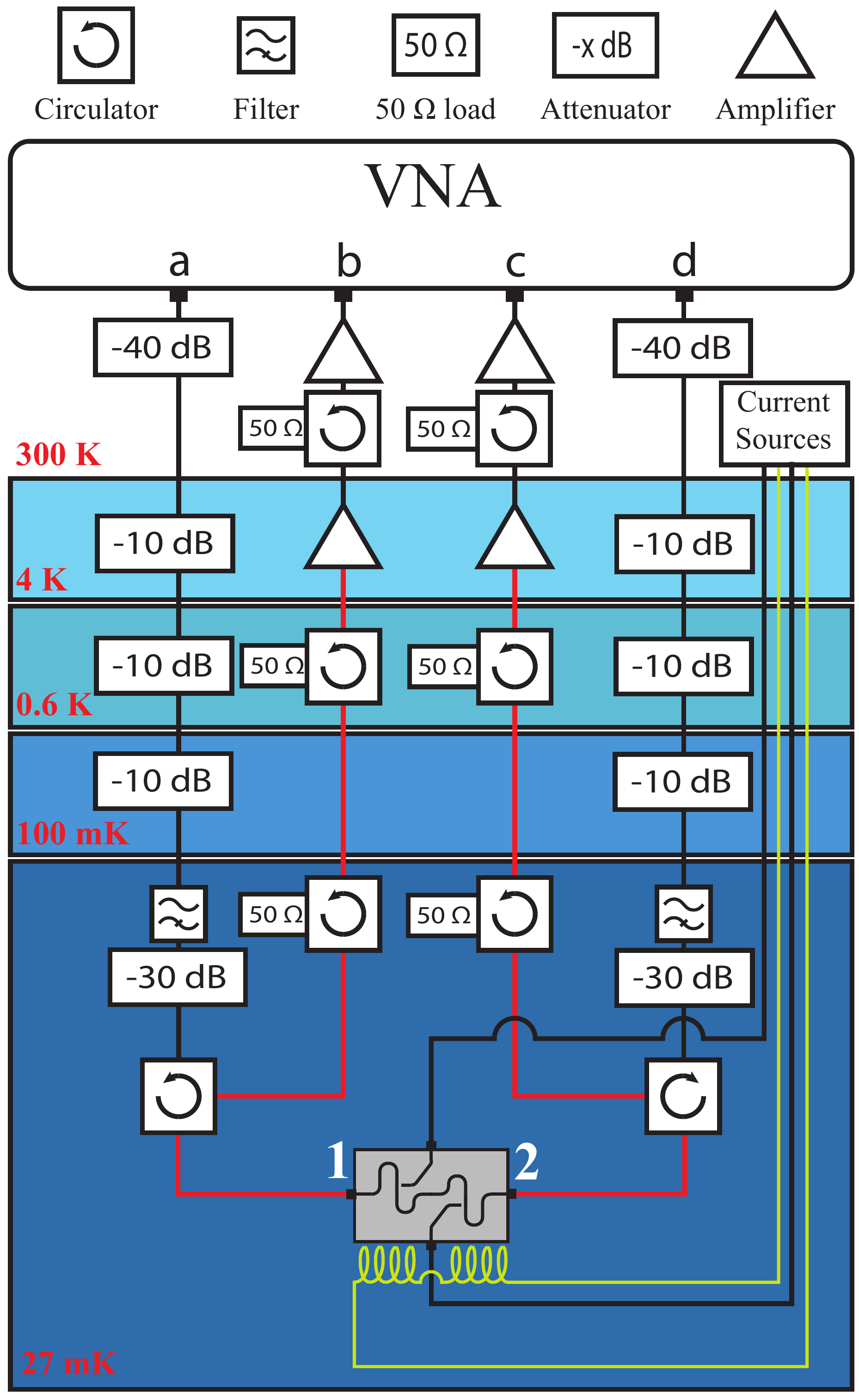}
\caption{Cryogenic setup. We apply a probe signal with a VNA (port a or d) through heavily attenuated input lines. The attenuation is distributed over multiple temperature stages of the cryostat to decrease the heat load on the system and to shield the sample from high-temperature noise. All output lines are made from superconducting niobium-titanium coaxial cables (red). In the output path, cryogenic circulators isolate the sample from high-temperature noise. The output signal is amplified with cryogenic and room temperature high-frequency amplifiers and then detected by the VNA (port b or c). The on-chip antennas and the external coil (yellow) are connected to current sources via twisted-pair wires. We label the feed line of the sample connected to the input port a by '1' and that connected to the port d by '2'.}
\label{fig:cryosetup}
\end{figure}

We mount the sample to the base plate of a dilution refrigerator with a base temperature of \SI{27}{mK} and apply a probe signal with a vector network analyzer (VNA) (see Fig.~\ref{fig:cryosetup}). The output signal is amplified with a cryogenic and a room temperature high-frequency amplifier and then detected by the VNA.

We can tune the critical current of our dc SQUIDS either by means of an external coil or via T-shaped on-chip antennas [see Fig.~\ref{fig:samplepic}(e)]. Although the antennas could guide microwave signals in future experiments, they are used to apply a quasi-dc flux bias to the SQUIDs in this work. The additional external coil facilitates the operation at flux bias values, where the currents through the antennas would start to introduce heating effects due to a breakdown of superconductivity. The external coil simultaneously tunes both SQUIDs, while the antennas are designed to individually access only one of them. However, we observe non-negligible crosstalk of each antenna to the other SQUID, which we have to account for in our experiments. 

\section{Nonlinearity from the circuit model}

In the scope of a quantum simulation experiment, it is vital to know the full parameter set of the underlying circuit in order to precisely predict its behavior. For a system of nonlinear resonators, the nonlinearity is of key interest. We therefore implement two ways to experimentally determine the nonlinearity of our system. In the following, we present a circuit model accurately reproducing our data (see Sec.~\ref{sec:circuitmodel}) and allowing us to calculate the nonlinearity (see Sec.~\ref{sec:nonlincircuit}). In Sec.~\ref{ch:Duff}, we compare these results to a direct measurement of the nonlinearity based on the power-dependent response of the resonators.

\subsection{Circuit model with competing signal path}
\label{sec:circuitmodel}

%
\begin{figure}
\centering
\includegraphics[width = 0.45\textwidth]{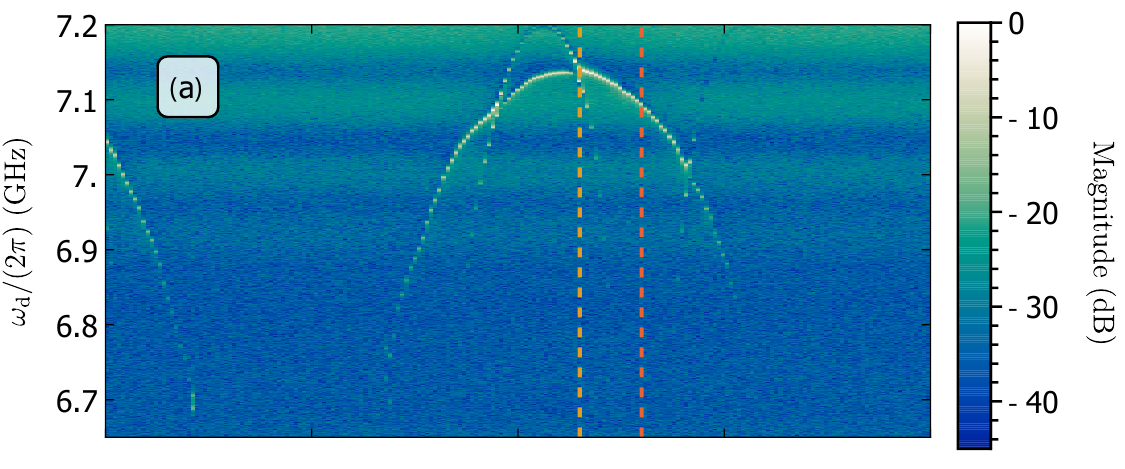}
\includegraphics[width = 0.45\textwidth]{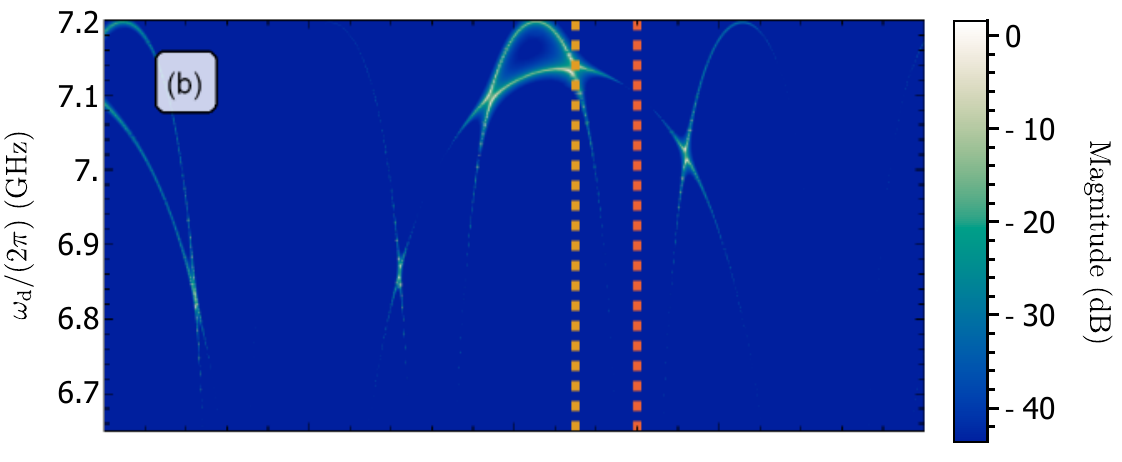}
\includegraphics[width = 0.45\textwidth]{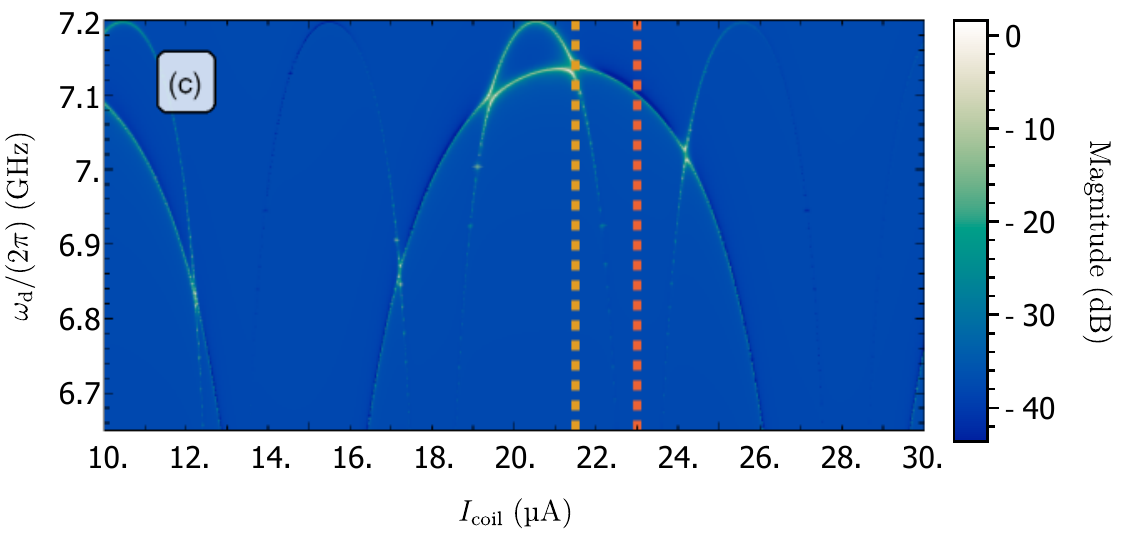}
\caption{Normalized transmission magnitude [$S_\text{ca}$ (VNA ports) or $S_{21}$ (sample ports)] through the sample as a function of the coil current and the probe frequency. (a) Measurement using the setup shown in Fig.~\ref{fig:cryosetup} with an input power of $P_\text{in} = \SI{-30}{dBm}$. (b) Simulation only considering the resonators and coupling capacitors. (c) Simulation taking into account a parasitic path. Data profiles at working points indicated by the dashed lines are displayed in Fig.~\ref{fig:simcuts}. Parameters used in the simulations are shown in Tab.~\ref{tab:respara} and Tab.~\ref{tab:paraspara}.}
\label{fig:sim}
\end{figure}

\begin{figure}
\centering
\includegraphics[width = 0.45\textwidth]{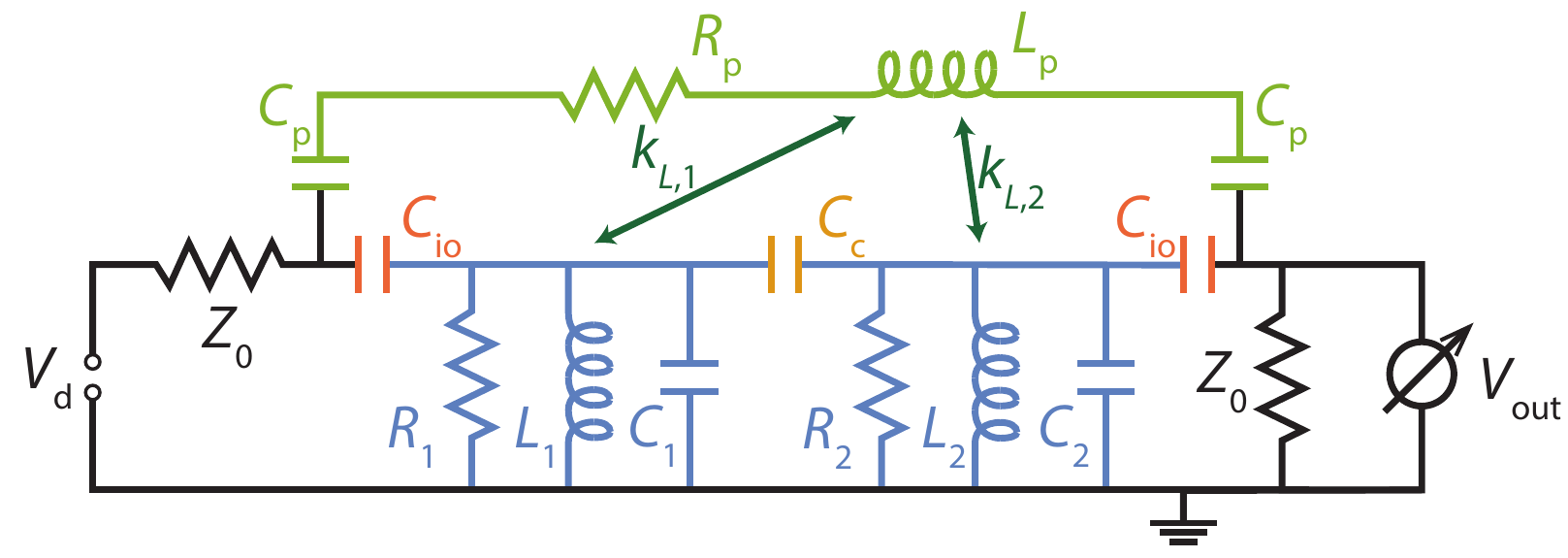}
\caption{Circuit model of the two resonator system. Resonators are shown in blue along with their coupling (yellow)  and input/output (red) capacitors. The parasitic path is shown in green and input and output lines in black. The path is capacitively coupled to the input and output lines and inductively coupled to the resonators with the strength quantified by the coupling constants $k_{L,i}$. Microwave drive and measurement signal are denoted by $V_\text{d}$ and $V_\text{out}$, respectively. Here, the contribution of the SQUIDs is included in the effective capacitance $C_{i}$ and inductance $L_{i}$ of the resonators.}
\label{fig:circmodel}
\end{figure}

\begin{figure}
\centering
\includegraphics[width = 0.45\textwidth]{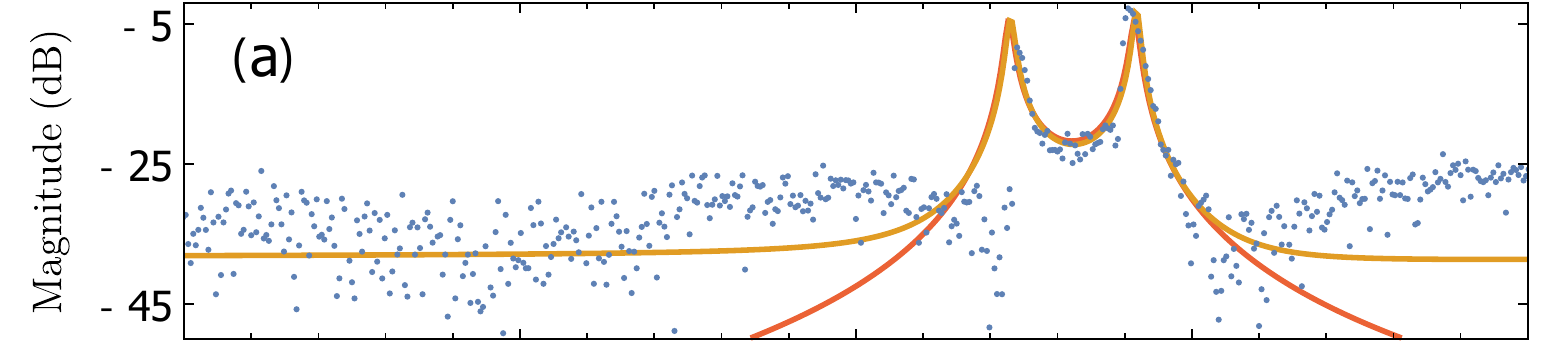}
\includegraphics[width = 0.45\textwidth]{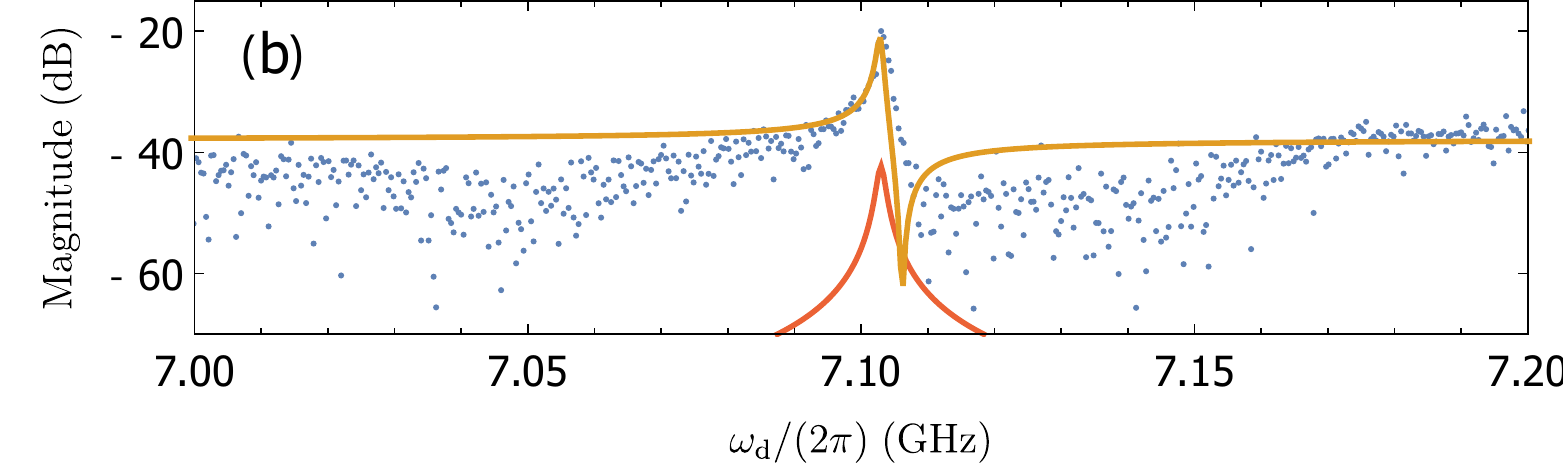}
\caption{Measured transmission magnitude  
$S_\text{ca}$ or $S_{21}$ (blue dots) as a function of the probe frequency for two fixed coil currents with input power $P_\text{in} = \SI{-30}{dBm}$. (a) Both resonators in resonance, $I_\text{coil}=\SI{21.5}{\micro\ampere}$. Yellow line: fit based on parasitic path model. Orange line: fit based on simple model. (b) Resonator 2 far detuned ($\omega_{r,2} = \SI{5.7}{\giga\hertz}$), $I_\text{coil}=\SI{23.0}{\micro\ampere}$. Yellow line: fit based on parasitic path model. Orange line: fit based on simple model.}
\label{fig:simcuts}
\end{figure}

In the experiment, we measure the transmission $S_\text{ca}$ or $S_{21}$ through our two-resonator system as a function of the current $I_\text{coil}$ through the coil and of the frequency $\omega_\text{d}/2\pi$ of the applied microwave drive. The result is shown in Fig.~\ref{fig:sim}(a). As expected, we observe periodic modulations of the two resonance frequencies of the coupled resonators. The maximum resonance frequencies of the two resonators differ by approximately \SI{50}{\mega\hertz} due to inaccuracies in the junction fabrication. For our sample, this amounts to a difference between the critical currents of the SQUIDs of roughly \SI{15}{\percent}. A spread in the critical currents of up to \SI{20}{\percent} is not unusual for our junction process~\cite{Wang2015}. Resonator 1 couples less strongly to the external magnetic field than resonator 2. This effect cannot be attributed to fabrication inaccuracies because of the relatively large size of the SQUID loop. Instead, we suspect strong local fields caused by asymmetric current flow across the superconducting ground plane.

In order to extract the circuits parameters, we first simulate the response of the sample with a simple circuit model taking into account only the resonators and coupling capacitors (see Fig.~\ref{fig:circmodel}). Details of the simulation can be found in App.~\ref{app:model}.
The simple circuit model predicts an increased transmission in regions where the resonators are close to resonance and a strongly suppressed transmission elsewhere [see Fig.~\ref{fig:simcuts}(a)]. 
Comparing the experimental data to the results of the simulation, we find good agreement when the two resonators are close to resonance with each other, but significant differences otherwise. For example, in the experiment, there is a clear transmission signal of resonator 1 even if resonator 2 is far detuned. The model, on the other hands, predicts a strong damping of the resonance of resonator 2 in this regime.
These observations become even more apparent when we look at the transmission signal for certain fixed coil currents. As shown by Fig.~\ref{fig:simcuts}(a), the simple model can reproduce the measured resonances qualitatively well, despite the fact that the measurement shows a larger background signal.
In contrast, Fig.~\ref{fig:simcuts}(b) shows that, when the two resonators are far detuned, the transmission through the system is predicted to be strongly damped over the whole frequency range. Even on resonance the measured peak is approximately \SI{15}{\decibel} higher than predicted.

We can account for these deviations by introducing a generic environment in form of a parasitic signal path (see Fig.~\ref{fig:circmodel}). This parasitic path consists of a series connection of resistive, inductive, and capacitive elements. They are coupled capacitively to the input and output lines and also inductively to the resonators. As shown in Fig. \ref{fig:sim}, the model that includes this parasitic path (see App.~\ref{app:model} for detailed calculations) reproduces the experimental data very well over the whole frequency range. The reason for this significant improvement is the fact that the parasitic path opens up additional transmission channels for the system. First of all, the signal can be directly coupled into the parasitic path via the input line and then be transmitted to the output line leading to an increased constant background even if both resonators are far detuned from the input frequency. Secondly, if the signal is in resonance with resonator 1, it can enter the resonator and then couple inductively to the parasitic path. This can be seen by the increased signal at the resonance frequency of resonator 1 even where resonator 2 is detuned.

Turning back to the frequency-dependent transmission at fixed coil current values, we can clearly distinguish between regions, where the path through the resonator system dominates, and regions, where the parasitic path plays a crucial role. In Fig.~\ref{fig:simcuts}(a), the two resonators have similar resonance frequencies and therefore transmit most of the signal to the output port. Hence, the parasitic path does not contribute to the shape of the resonances. Away from the resonances, the broadband nature of the parasitic path allows for an increased transmission background as it is observed in the measurements. When the two resonators are far detuned, the presence of the parasitic path also changes the qualitative shape of the resonance, making it Fano-like. While the peaks are symmetric in the simple model at all times [see Fig.~\ref{fig:simcuts}(b)], the  parasitic path model matches the peak-dip feature of the measurement [Fig.~\ref{fig:simcuts}(b)].
In summary, based on the parasitic path model, we obtain a realistic set of parameters for each resonator (see Tab.~\ref{tab:respara} and Tab.~\ref{tab:paraspara}). Additionally, we calculate the capacitance per unit length $C_0 = \SI{0.18}{\nano\farad\per\metre}$ and the inductance per unit length $L_0 = \SI{0.44}{\micro\henry\per\metre}$ of the resonators from their bare resonance frequency. Our simulations show that both capacitive and inductive coupling are necessary in order to correctly model the experimental results. For definitions and further explanations regarding these parameters, please refer to App.~\ref{app:model}.

\begin{table}
\centering
\begin{tabular}{c|c|c|c|c|c}
 & $I_\text{c} (\si{\micro\ampere})$ & $\delta\Phi(\Phi_0)$ & $\Delta\Phi/\Delta I_\text{coil} (\Phi_0\si{\per\micro\ampere})$ & $d$ & $k_{L} (10^{-3})$ \\ 
\hline 
Res 1 & 1.56 & -0.39 & 0.076 & 0.13 & 8\\ 
\hline 
Res 2 & 1.80 & 0.30 & 0.180 & 0.13 & 0.75\\ 
\end{tabular} 
\caption{Resonator parameters extracted from the circuit model. For each resonator, we show the total critical current $I_\text{c} = I_\text{c1} + I_\text{c2}$ of the SQUID, the zero current offset $\delta\Phi$ of the flux through the SQUID loop, the flux change $\Delta\Phi$ per applied coil current $\Delta I_\text{coil}$, the SQUID asymmetry parameter $d$, and the inductive coupling constant $k_{L}$ to the parasitic path. For better readability, the index i representing each resonator has been omitted in this table.}
\label{tab:respara}
\end{table}

\begin{table}
\centering
\begin{tabular}{c|c|c} 
$C_\text{p} (\si{\femto\farad})$ & $R_\text{p} (\si{\ohm})$ & $L_\text{p} (\si{\nano\henry})$ \\ 
\hline 
6.2 & 8000 & 133\\  
\end{tabular} 
\caption{Parameters of the parasitic path (see Fig.~\ref{fig:circmodel}). Here, we show the capacitance $C_\text{p}$, resistance $R_\text{p}$ and inductance $L_\text{p}$ of the parasitic path.}
\label{tab:paraspara}
\end{table}

\subsection{Calculation of the nonlinearity from the circuit model}
\label{sec:nonlincircuit}
\begin{figure}
\centering
\includegraphics[width = 0.45\textwidth]{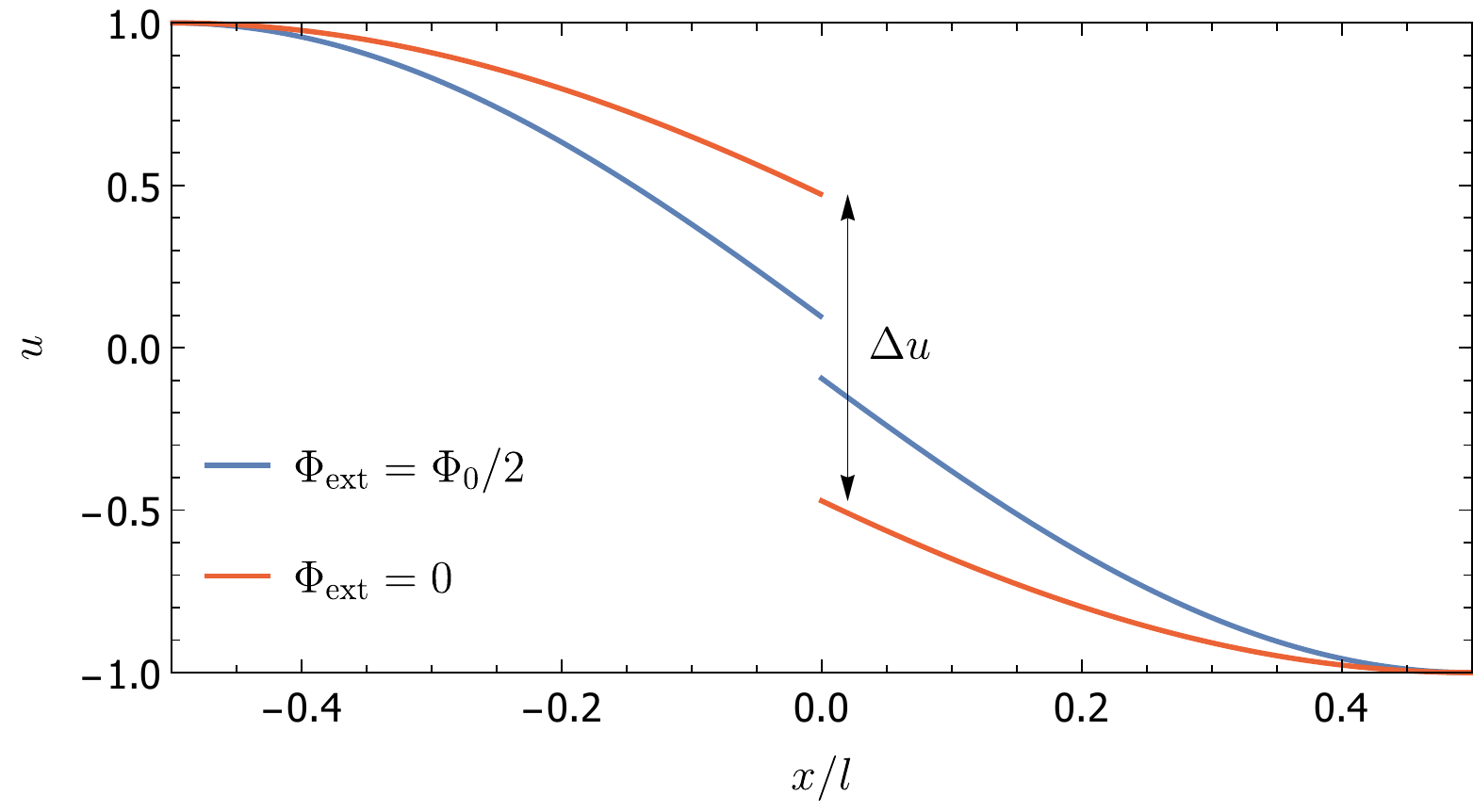}
\caption{Dimensionless envelope $u$ of the first spatial voltage mode at $\Phi_\mathrm{ext} = 0$ and $\Phi_\mathrm{ext} = \Phi_0/2$ of a resonator of length $l$ with a SQUID at position $x = 0$. The difference of the spatial mode across the SQUID, $\Delta u$, is a direct measure of the nonlinearity of the resonator.}
\label{fig:u1}
\end{figure}
Using the parameters extracted from the parasitic-path model discussed in the previous section, we can estimate the nonlinearity $U$ created by the SQUID for a single half-wavelength coplanar waveguide resonator following Ref.~\cite{Bourassa2012} via the relation
\begin{align}
\label{eq:U}
U = -\frac{e^2}{2\hbar L_\text{J}}\frac{L\Delta u ^4}{C}.
\end{align}
Here, the parameters $C$ and $L$ are the effective capacitance and inductance of the resonator including the contribution from the SQUID. We numerically calculate $\Delta u$, which is the difference of the dimensionless spatial voltage mode envelope $u$ of the first resonator mode across the point-like SQUID (see Fig.~\ref{fig:u1} for details). For a detailed derivation, see App.~\ref{app:model}.

\begin{figure}
\centering
\includegraphics[width = 0.45\textwidth]{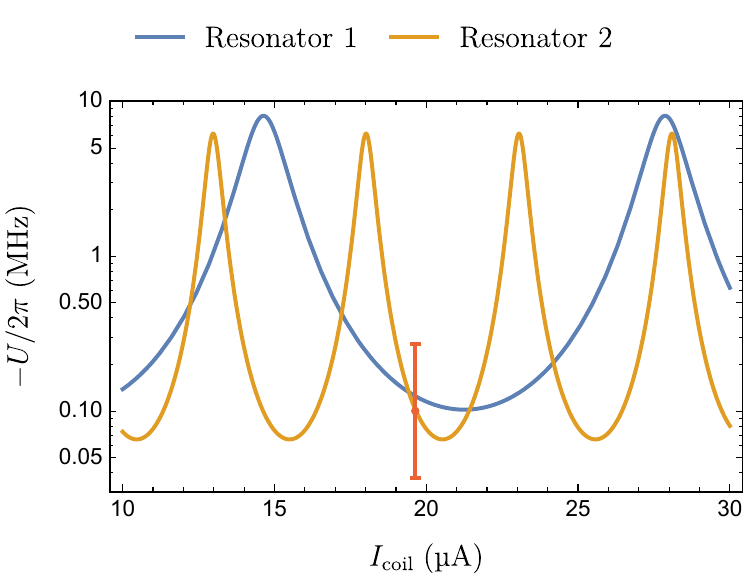}
\caption{The nonlinearity $U$ as a function of the current flowing through the external coil. Solid lines represent calculations based on parameters extracted from transmission data and parasitic-path model for resonator 1 (blue) and resonator 2 (yellow). The red dot is the result of a direct measurement for resonator 2 described in Sec.~\ref{ch:Duff}.}
\label{fig:logU}
\end{figure}

For our two resonators, the dependence of $U$ as a function of $I_\text{coil}$ is shown in Fig.~\ref{fig:logU}. The absolute value of the nonlinearity of resonator 1 (2) can be tuned between a minimum of \SI{0.1}{\mega\hertz} (\SI{0.06}{\mega\hertz}) and a maximum of \SI{8.0}{\mega\hertz} (\SI{6.1}{\mega\hertz}).  Due to the difference in the maximal critical currents of the two dc SQUIDs, the tuning ranges for the nonlinearity differ slightly. Nonetheless, they extend over almost two orders of magnitude. This fact allows us to set the nonlinearity $U$ \textit{in situ} between values well below the resonator-resonator coupling rate $J = \SI[separate-uncertainty=true,multi-part-units=single]{7.6\pm0.3}{MHz}$ and values well above $J$ by changing the magnetic bias fields. $J$ is extracted from the level splitting at the frequency degeneracy point of the two resonators. For more details, see App.~\ref{sec:J}.

\section{Nonlinearity from a direct power-dependent measurement}
In addition to the values extracted from the circuit model in the previous section, we present a direct measurement of the nonlinearity of resonator 2. Specifically, we exploit the response of the resonance frequency as a function of the input power. The relevant parameter to determine $U$ is the actual power circulating inside the resonator. Therefore, we first have to extract the external coupling strength between resonator and transmission line to convert the applied power to the field strength inside the resonator. 
\subsection{Quality factor}
\label{sec:QualityFactor}

In order to extract the external quality factor of our resonators, we use an input-output formalism~\cite{Walls1995} and fit the result to the measured reflection signal of the two-resonator chain. 
In the limit where resonator 1 is far detuned, we find a dependency of the scattering parameter $S_{22}$ on the quality factors of resonator 2,

\begin{align}
\label{eq:Qext1}
S_{22} \approx 1 - \frac{2Q_{\ell,2}/Q_\text{ext,2}}{1 -2iQ_{\ell,2}\left(\omega_\text{r,2} - \omega_\text{d}\right)}.
\end{align} 
Here, we have used the loaded quality factor $Q_\text{l,2}$, the external quality factor $Q_\text{ext,2}$ and the resonance frequency $\omega_\text{r,2}$, each of resonator 2. The parameter $\omega_\text{d}$ denotes the angular frequency of the driving field.
On resonance of the second resonator, $\omega_\text{r,2} - \omega_\text{d} = 0$, Eq.~(\ref{eq:Qext1}) further simplifies to
\begin{align}
\label{eq:Qext}
S_{22} \approx 1 - \frac{2Q_{\ell,2}}{Q_\text{ext,2}}.
\end{align} 

\begin{figure}
\centering
\includegraphics[width = 0.45\textwidth]{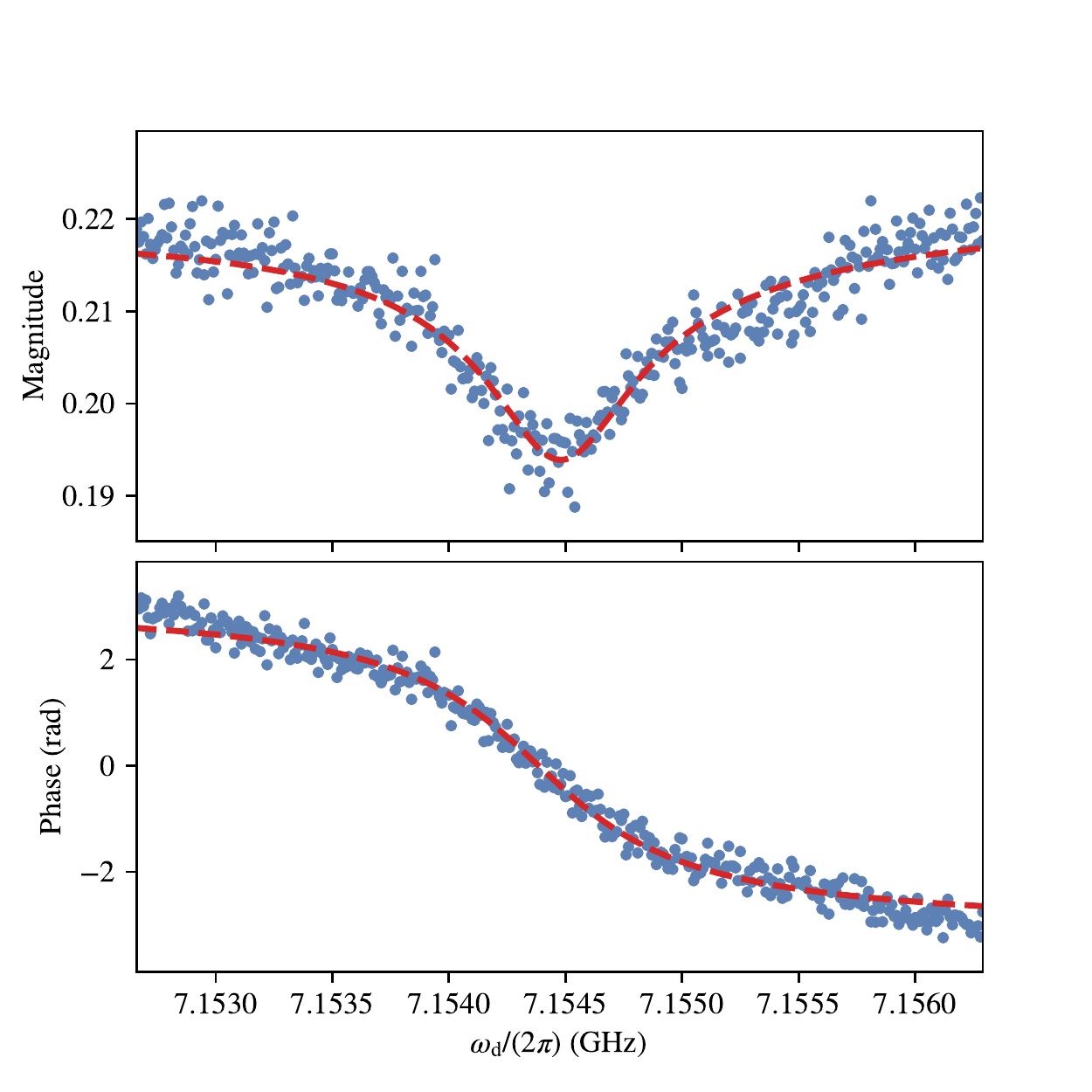}
\caption{Fit (red dashed line) to the reflection data $S_\text{cd}$ or $S_{22}$ (blue dots) at the maximum frequency of resonator 2 when resonator 1 is far detuned.}
\label{fig:InOutFit}
\end{figure}

Further information on the derivation of these equations can be found in App.~\ref{app:qfactor}. We first fit the predicted phase dependence, $\theta = \theta_0 + 2 \arctan(2Q_{\ell,2}(1-\omega/\omega_\text{r,2}))$, to the measured scattering parameter data to extract the loaded quality factor. Then, we fit Eq.~(\ref{eq:Qext1}) to the magnitude and use Eq.~(\ref{eq:Qext}) to determine the external quality factor (see Fig.~\ref{fig:InOutFit}).
For resonator 2, we obtain an external quality factor of $Q_\text{ext,2} = 1.35\times10^5$.

\subsection{Nonlinearity from power-dependent resonance amplitude}
\label{ch:Duff}
In order to get a relation between the directly measurable output voltage of our system and the nonlinearity, we start with the equation of motion for a single resonator driven with strength $F_0$. It can be written in terms of the flux $\Psi = \int V(x,t)\text{d}t$, where $V(x,t)$ is the internal voltage of the resonator,

\begin{equation}
\frac{\Psi}{L} + C\ddot{\Psi} + \frac{\dot{\Psi}}{R} + \beta\Psi^3 = F_0 e^{i\omega t}.
\end{equation}
Here,
\begin{equation}
\label{eq:betamod}
\beta = -\frac{1}{24}\left(\frac{2\pi}{\Phi_0}\right)^2\frac{\Delta u^4}{L_\text{J}}
\end{equation}
is the prefactor of the nonlinear term due to the tunable Josephson junction formed by the SQUID. The parameter $\beta$ depends on the SQUID inductance $L_\text{J}$ and the drop $\Delta u$ in the spatial voltage mode across the SQUID (see also Fig.~\ref{fig:u1}). Obviously, the prefactor $\beta$ is a direct measure for the nonlinearity of the system.

For the Duffing-like equation of motion, we can show that the maximum amplitude $a$ of the mode is inversely proportional to the nonlinearity for small deviations from the unperturbed resonance frequency $\omega_0$~\cite{Nayfeh1983}
\begin{equation}
\left|\Psi\right|^2= a^2  = \frac{8}{3}\frac{\omega_0 C}{\beta}\left(\omega-\omega_0\right).
\end{equation}
For the experimental output voltage we derive (App.~\ref{app:NonLin})
\begin{equation}
\label{eq:VtoFPwer}
V_\text{out}^2 = \frac{8}{3}\frac{\omega_0 C}{\beta}\left(\omega_0-\omega\right)\frac{Z_0\omega}{2Q_\text{ext}L}G,
\end{equation}
with $Z_0 = \SI{50}{\ohm}$ being the characteristic impedance of the circuit and $G$ the gain of the amplification chain. 

\begin{figure}
\centering
\includegraphics[width = 0.45\textwidth]{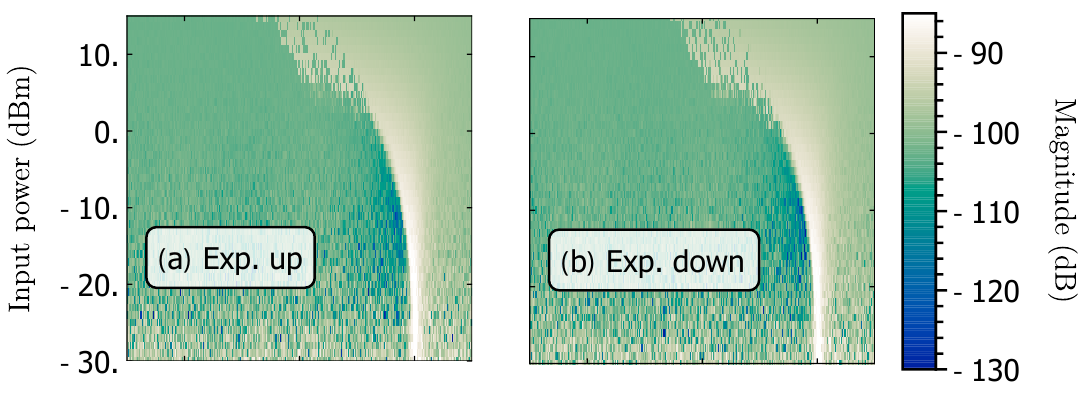}
\includegraphics[width = 0.45\textwidth]{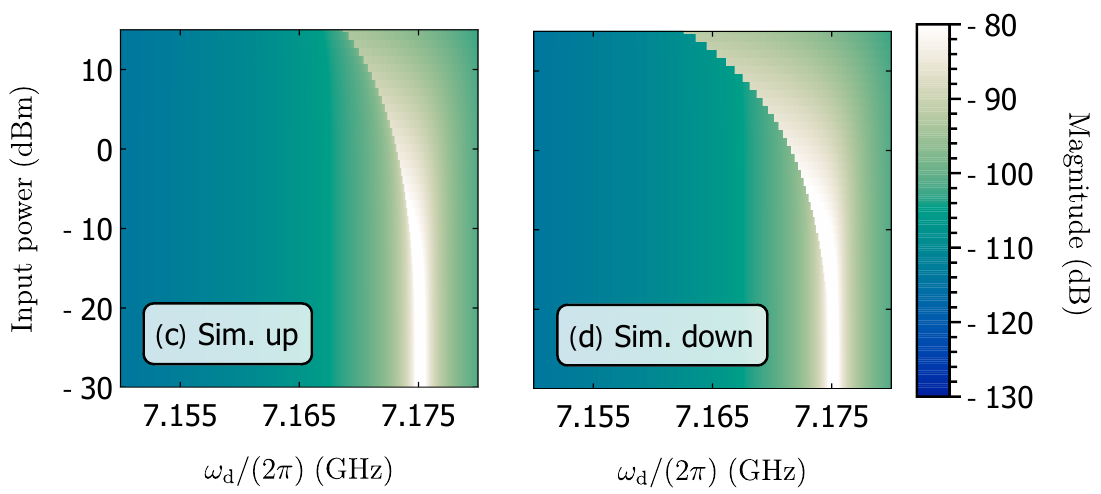}
\caption{Frequency-dependent transmission measurements $S_\text{ca}$ or $S_{21}$ near the maximum frequency of resonator 2 as a function of the VNA output power. For this experiment, resonator 2 is at \SI{7.1}{\giga\hertz}. (a), (b) Experimental data and  (c), (d) simulation using parameters extracted from the parasitic-path model. Frequency up(down)-sweeps are labelled with "up" ("down"). The simulation has been calibrated with the input attenuation and output amplification measured for our setup (for details, see App.~\ref{app:PNCF}).}
\label{fig:expPowerSweep}
\end{figure}


In order to obtain information on $\beta$, we perform power-dependent measurements of the transmission through the tunable resonator system near the maximum resonance frequency of resonator 2. The corresponding data is shown in Fig.~\ref{fig:expPowerSweep}. As expected  for a softening nonlinearity $U<0$, the resonance frequency decreases with increasing power. We adjust the previous circuit model to include the purely nonlinear part of the equation of motion, $\beta\Psi^3$, and model the power dependence of the system. The current is adjusted by a nonlinear perturbation of the linear current $I$
\begin{equation}
I_\text{NL} = I_i -\frac{\beta}{i\omega_\text{d}^3}V_i^3.
\end{equation}
Here, we use $\Psi = \int V\text{d}t = V/(i\omega)$ to calculate the perturbation. Comparing the frequency up-sweep to the down-sweep, both for the measurement and simulation, we can observe a region of bistability for high input powers (starting at roughly \SI{0}{dBm}). While the simulation predicts that the system is in different but stable states during sweeping up and down, the measurement shows jumps between the high and the low transmission state for both the up and down sweep.
We find a good agreement for the upper and lower frequency bound of the bistability region between experiment and model (see Fig.~\ref{fig:expPowerSweep}).

\begin{figure}
\centering

\subfigure{\includegraphics[width = 0.45\textwidth]{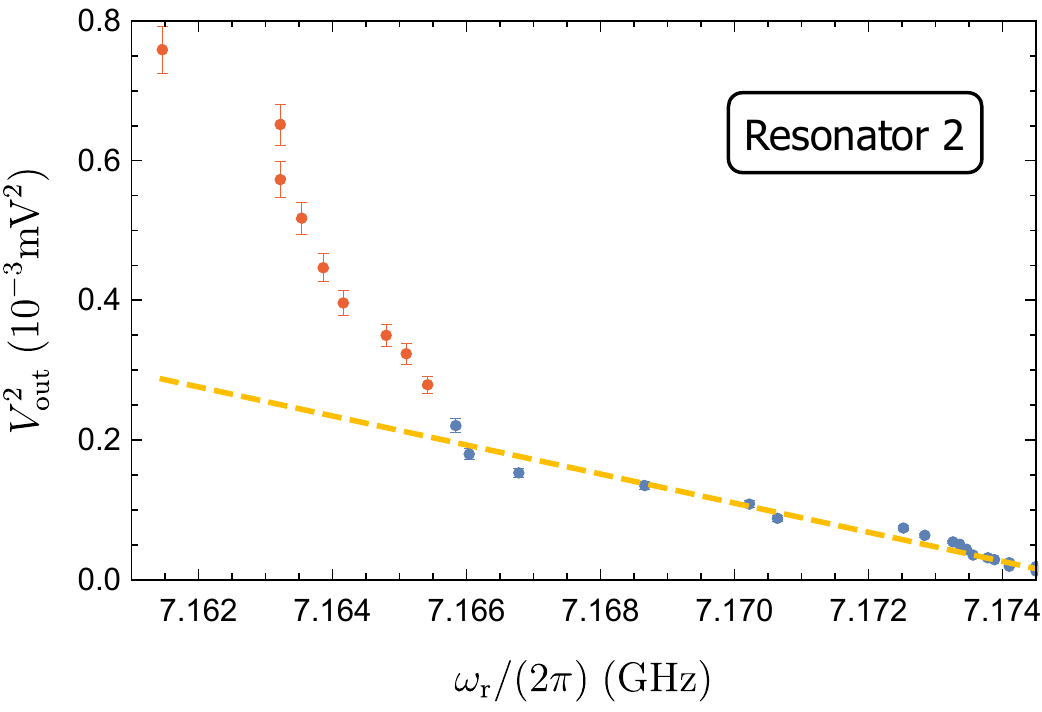}}
\caption{Squared output voltage at the resonance frequency $\omega_\text{r}$ as a function of this frequency for resonator 2 (blue and red dots) and the respective fit (dashed yellow line) of Eq.~(\ref{eq:VtoFPwer}) to data points of low input power (blue dots). If not shown, the error bars are smaller than the symbol size. We estimate the uncertainty of each frequency point to be $\pm\SI{2}{\mega\hertz}$.}
\label{fig:V2F}
\end{figure}

In order to calculate the nonlinearity, we plot the square of the output voltage against the effective resonance	 frequency of the resonator, which is given by the maximum amplitude of the resonator response. As expected for a system with softening nonlinearity, the output voltage increases with decreasing resonance frequency. For small deviations from the unperturbed frequency, this increase is expected to be linear [see Eq.~(\ref{eq:VtoFPwer})]. The experimental data, taken from the transmission measurement shown in Fig.~\ref{fig:expPowerSweep}, indeed shows this linear increase for low input power (see Fig.~\ref{fig:V2F}).
We use $C= C_2$ and $L= L_2$, i.e., the effective capacitance and inductance of resonator 2 from the parasitic-path model. As we look at the nonlinearity of resonator 2 in a transmission measurement, where we apply a signal at the input of resonator 1 and measure at the output of resonator 2, the measured output signal can be directly related to the voltage inside resonator 2 via the external quality factor determined in Sec.~\ref{sec:QualityFactor}.
From a photon number calibration measurement (see App.~\ref{app:PNCF}), we estimate that the gain of our amplification chain is $G = \SI[separate-uncertainty=true,multi-part-units=single]{38\pm4}{dB}$. The uncertainty of the following results is calculated using error propagation, where the main contribution is due to the systematic uncertainty of the gain $G$ Here, we choose a rather pessimistic estimate. Still, the error bar is only \SI{4}{\percent} with respect to the total tuning range. As the gain uncertainty is estimated in decibel, the error bars are asymmetric in linear units.
A numerical fit of Eq.~(\ref{eq:VtoFPwer}) to the squared output voltage as a function of the resonance frequency in the low power region (Fig.~\ref{fig:V2F}) yields $\beta = 1.97\substack{+3.03 \\ -1.19}$\,\si{\ampere\per\volt\cubed\second\per\cubed}, which we can directly relate to the nonlinearity $U$ via Eq.~(\ref{eq:U}) and Eq.~(\ref{eq:betamod}). For resonator 2, we get $U_\text{res2} = 0.10\substack{+0.16 \\ -0.06}$\,\si{\mega\hertz}.

We find that the predictions of our theoretical model for the nonlinearity agree well with our measurement within its uncertainty (see Fig.~\ref{fig:logU}).

\section{Conclusion}
We have investigated a superconducting circuit consisting of two tunable and coupled nonlinear resonators. The nonlinearity is induced by a dc SQUID galvanically coupled to each resonator. The system can be fully controlled by means of an external coil and two on-chip antennas, allowing us to tune the nonlinearity by roughly two orders of magnitude. The nonlinearity of resonator 1 (2) can be tuned between a minimum of \SI{0.1}{\mega\hertz} (\SI{0.06}{\mega\hertz}) and a maximum of \SI{8.0}{\mega\hertz} (\SI{6.1}{\mega\hertz}). We have shown that we are able to model the response of our two-resonator system with an equivalent circuit including an additional signal path. In this way, we can reliably simulate the experimentally obtained transmission data. We have confirmed the nonlinearity extracted from the circuit model by means of direct, power-dependent transmission measurements. As a result of the demonstrated control of the nonlinearity and the understanding of the environment, the studied system is a promising candidate for quantum simulations of a driven-dissipative Bose-Hubbard physics.
We acknowledge support by the Elite Network of Bavaria through the program ExQM, the EU Quantum Flagship project QMiCS (Grant No. 820505), and the German Federal Ministry of Education and Research (BMBF) via the project QUARATE (Grant No. 13N15380).


\addtocontents{toc}{\protect\vspace{1cm}}
\addcontentsline{toc}{chapter}{Bibliography}

\bibliographystyle{bibstyle}

\appendix

\section{Derivation of the circuit model with a competing path}
\label{app:model}
Here, we present the full circuit model for our two resonator system including a competing path. The standard model without this additional path can be obtained if we neglect all parts containing the parasitic path. For the driving voltage ($V_\text{d}$) and output voltage ($V_\text{out}$), we find
\begin{align}
V_\text{d} &= Z_0 (I_1 + I_\text{p}) + V_\text{x}\\
V_\text{out} &= Z_0 (I_3 + I_\text{p}),
\end{align}
with $I_1$ ($I_3$) being the current flowing into (out of) the two resonator system. $V_\text{x}$ is the voltage drop across the impedance $Z_0$ of the input cable. 

We can describe the currents $I_i$ flowing in our system using the voltages $V_i$ at the in- and output capacitors (capacitance $C_\text{io}$), the coupling capacitor (capacitance $C_\text{c}$) and the drive frequency $\omega_d$. The index $i$ denotes the capacitor starting from driving side and $j$ the resonator number. 
\begin{align}
I_1 &= (V_\text{x}-V_1)i\omega_d C_\text{io}\\
I_2 &= (-V_1-V_2)i\omega_d C_\text{c}\\
I_3 &= (-V_2-V_\text{out})i\omega_d C_\text{io}.
\end{align}
As we are only looking at the first voltage modes, the signs of $V_1$ and $V_2$ change at their second appearance.

In each resonator, the inflowing and outflowing currents have to be the same due to current conservation:
\begin{align}
I_1 = (i\omega_d C_1+\frac{1}{R_1}+\frac{1}{i\omega_d L_1})V_1 - k_{L,1}\frac{\sqrt{L_1L_\text{p}}}{L_1}I_\text{p}-I_2,\\
I_2= (i\omega_d C_2+\frac{1}{R_2}+\frac{1}{i\omega_d L_2})V_2 - k_{L,2}\frac{\sqrt{L_2L_\text{p}}}{L_2}I_\text{p}-I_3.
\end{align}

Here, $C_j$, $L_j$ and $R_j$ are the capacitance, inductance and resistance of resonator $j = 1,2$ and $k_{L,j}$ is its inductive coupling strength to the parasitic path. $L_\text{p}$, $C_\text{p}$, $R_\text{p}$ and $I_\text{p}$ are the inductance, capacitance and resistance of the parasitic path and the current flowing through it.
Additionally we consider the voltage drops inside the parasitic path:
\begin{align}
\begin{aligned}
\frac{1}{i\omega_d C_\text{p}}I_\text{p} -& i\omega_d k_{L,1} \sqrt{L_1L_\text{p}}\frac{V_1}{i\omega_dL_1}+
i\omega_d k_{L,2} \sqrt{L_2L_\text{p}}\frac{V_2}{i\omega_dL_2}\\ 
+&i\omega_d L_\text{p} I_\text{p} + R_\text{p}I_\text{p} +\frac{1}{i\omega_d C_\text{p}} I_\text{p} = V_\text{x} - V_\text{out}.
\end{aligned}
\end{align}

To include the tunability of the circuit introduced by the dc SQUIDs, we calculate $C_i$ and $L_i$ as functions of the external flux $\Phi_\text{ext}$ penetrating the SQUID loops:

\begin{align}
C_i = C_0 \int_{-L/2}^{+L/2}  u\left(x, k,\Phi_\text{ext}\right)^2\text{dx} + C_i \Delta u\left(k,\Phi_\text{ext}\right)^2\\
\begin{aligned}
L_i &= \left(\frac{1}{L_0} \int_{-L/2}^{+L/2}\left(\delta_\text{x} u\left(x,k,\Phi_\text{ext}\right)^2\right)\text{dx} \right.\\
&\quad\left. + \frac{1}{L_i\left(\Phi_\text{ext}\right)}\Delta u\left(k,\Phi_\text{ext}\right)^2\right)^{-1}.
\end{aligned}
\end{align}

With a standard ansatz for the dimensionless envelopes of the spatial mode functions $u_i$ of the resonator on the left and right side of the SQUID
\begin{align}
u_{i,\ell} &= A_\text{l}\cos\left(k\left(x+L/2\right)\right)\\
u_{i,\text{r}} &= A_\text{r}\cos\left(k\left(x-L/2\right)\right),
\end{align}
we can calculate $u(k)$ and the jump of the mode function $\Delta u(k)$ at the position of the SQUID after extracting the wave vector $k$ from 
\begin{align}
k^{\text{odd}} \tan\left( k^{\text{odd}}\frac{L}{2}\right)  = - \frac{\text{Z}_0}{\nu}C_i\left(\omega_\text{p}^2-\left(k^{\text{odd}}\nu\right)^2\right).
\label{equ:k}
\end{align}
As the plasma frequency $\omega_\text{p} = 1/(C_\text{J}L_\text{J})$ depends on the Josephson inductance $L_\text{J}$, which, in turn, depends on $\Phi_\text{ext}$, the full circuit model is flux dependent.
$L_\text{J}$ can be described by the supercurrent $I_\text{s}$ flowing through each SQUID loop
\begin{align}
I_\text{s} &= I_\text{c}\left|\cos\left(\pi\frac{\Phi_\text{ext}}{\Phi_0}\right)\right| \sqrt{1+d^2\tan\left(\pi\frac{\Phi_\text{ext}}{\Phi_0}\right)}
\end{align}
Here, $L_\text{J} = \frac{\Phi_0}{2\pi I_\text{s}}$ is the Josephson inductance and $d = \left(I_\text{c1}-I_\text{c2}\right)/\left(I_\text{c1}+I_\text{c2)}\right)$ the asymmetry parameter of the dc SQUID. $I_{\text{c}i}$ is the critical current of junction $i$ and $I_\text{c}$ the total critical current.

\section{Maximum of the frequency response function}
\label{app:NonLin}
In order to use the duffing equation of motion and its frequency response to model our experimental data, we have to transform the equations into experimentally accessible parameters. Therefore we use $\Psi = \int V\text{d}t = V/(i\omega)$ to transform 
\begin{equation}
\left|\Psi\right|^2= a^2  = \frac{8}{3}\frac{\omega_0 C}{\beta}\left(\omega-\omega_0\right)
\end{equation}
into 
\begin{equation}
\frac{V^2}{\omega^2} = \frac{8}{3}\frac{\omega_0 C}{\beta}\left(\omega-\omega_0\right).
\end{equation}
Here, $V$ is the voltage of the internal mode, which cannot be directly measured. We can use the external quality factor in order to relate the internal and external voltages~\cite{Pozar2012}
\begin{equation}
Q_\text{ext} = \omega \frac{W_\text{m}+W_\text{e}}{P_\text{loss}} = \omega \frac{\left|V_\text{int}\right|^2/(2L\omega^2)}{\left|V_\text{ext}\right|^2/Z_0}.
\end{equation}
We assume that, on resonance, the energy $W_\text{e}$ stored in the capacitance is equal to the energy $W_\text{m}$ stored in the inductance and that the power of an electrical signal in our waveguide can be written as $P = V^2/Z_0$.
We therefore obtain 
\begin{equation}
\frac{V_\text{int}^2}{\omega^2} = \frac{2V_\text{out}^2 Q_\text{ext}L}{Z_0\omega}.
\end{equation}
For the experimentally accessible voltage $V_\text{out}$, we obtain a similar dependence as for the internal flux field, but modified with an additional scaling factor $(Z_0\omega G)/(2Q_\text{ext}L)$
\begin{equation}
\label{eq:VtoFPwerApp}
V_\text{out}^2 = \frac{8}{3}\frac{\omega_0 C}{\beta}\left(\omega_0-\omega\right)\frac{Z_0\omega}{2Q_\text{ext}L}G.
\end{equation}
Here, we take the power gain $G$ in our experiment into account as we do not directly measure the output voltage of the resonators but a voltage after amplification (see Fig. \ref{fig:cryosetup}).

\section{Calculation of the external quality factor}
In App.~\ref{app:NonLin}, it is shown that, in order to extract the nonlinearity from a power dependent measurement, we need to know the external quality factor of the resonators. 
Here, we derive equations, that allow us to extract the external quality factor from a reflection measurement of our two resonator system.
\label{app:qfactor}
\begin{figure}
  \centering
  \includegraphics[width=0.7\columnwidth]{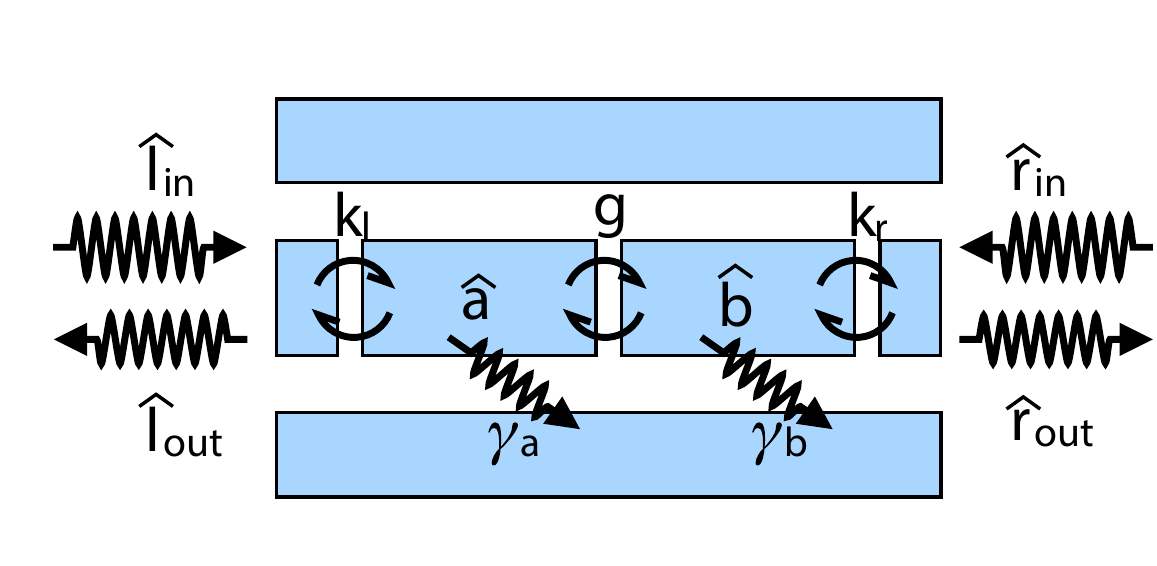}
  \caption{Quantum description of the system. The two intra-resonator fields, $a$ and $b$, are coupled with a coupling strength $g$, which are also coupled to two independent baths with strengths $\kappa_{\rm l}$ and $\kappa_{\rm r}$, respectively. From the two baths, one can identify the input and output fields as $l_{\rm in}$ and $l_{\rm out}$, or $r_{\rm in}$ and $r_{\rm out}$, depending on the boundary conditions. In addition, we denote the intrinsic losses of the two cavities as $\gamma_{\rm a}$ and $\gamma_b$, respectively. The hat symbols denoting quantum operators in the sketch are omitted in the text to simplify the notation.}
  \label{fig:quality_factor}
\end{figure}

First, we consider a system without intrinsic damping, of which the Hamiltonian reads \cite{Gardiner1985, Gardiner1993, Carmichael1993}
\begin{align}
	H_{\rm sys} &= \hbar\omega_{\rm a} a^{\dagger}a + \hbar\omega_{\rm b} b^{\dagger}b + \hbar g\left( a^{\dagger}b + ab^{\dagger} \right) \\
	H &= H_{\rm sys} + \hbar\int_{-\infty}^{+\infty} d\omega \left\{\omega l^{\dagger}\left(\omega\right)l\left(\omega\right) 
	+\right. \\ 
	&\quad\quad\quad\quad\quad\quad\quad\left. i\kappa_l\left(\omega\right) \left[l^{\dagger}\left(\omega\right)a - l\left(\omega\right)a^{\dagger} \right] \right\} \nonumber \\
	&~~~~~~~~~~
	+ \hbar\int_{-\infty}^{+\infty} d\omega \left\{\omega r^{\dagger}\left(\omega\right)r\left(\omega\right)+\right. \nonumber\\
	&\quad\quad\quad\quad\quad\quad\quad\left. i\kappa_l\left(\omega\right) \left[ r^{\dagger}\left(\omega\right)b - r\left(\omega\right)b^{\dagger} \right] \right\} \nonumber
\end{align}
Here, by convention, we define the specific type of coupling between the intra-resonator fields, $a$ for resonator 1 and $b$ for resonator 2, and the bath, $l\left(\omega\right)$ and $r\left(\omega\right)$, respectively, for the simplicity of derivation. Then, one can derive the following Heisenberg equations of motion for the field operators 
\begin{align}
	\dot{l}\left(\omega\right) &= -i\omega l\left(\omega\right) + \kappa_\ell\left(\omega\right)a, \label{eq:cl}\\
	\dot{a} &= -\frac{i}{\hbar}\left[a, H_{\rm sys}\right] - \int_{-\infty}^{+\infty} d\omega \kappa_\ell\left(\omega\right)l\left(\omega\right), \label{eq:a}\\
	\dot{r}\left(\omega\right) &= -i\omega r\left(\omega\right) + \kappa_r\left(\omega\right)b, \label{eq:cr}\\
	\dot{b} &= -\frac{i}{\hbar}\left[b, H_{\rm sys}\right] - \int_{-\infty}^{+\infty} d\omega \kappa_r\left(\omega\right)r\left(\omega\right). \label{eq:b}
\end{align}
We note that the above equations can be split into two groups, namely Eq.~\eqref{eq:cl}/\eqref{eq:a} and Eq.~\eqref{eq:cr}/\eqref{eq:b}, each of which is identical to the input-output formalism of a single system. Following the same procedure as in Ref.~\onlinecite{Gardiner1985}, we define the input fields
\begin{align}
	l_{\rm in} &= \frac{1}{\sqrt{2\pi}}\int_{-\infty}^{+\infty} d\omega e^{-i\omega t}l\left(\omega\right), \\
	r_{\rm in} &= \frac{1}{\sqrt{2\pi}}\int_{-\infty}^{+\infty} d\omega e^{-i\omega t}r\left(\omega\right).
\end{align}
The equations of the intra-resonator fields thus read
\begin{align}
	\dot{a} &= -i\omega_{\rm a}a - igb - \frac{\gamma_{\rm \ell} + \gamma_{\rm a}}{2}a -\sqrt{\gamma_{\rm \ell}}l_{\rm in}(t), \\
	\dot{b} &= -i\omega_{\rm b}b - iga - \frac{\gamma_{\rm r} + \gamma_{\rm b}}{2}b -\sqrt{\gamma_{\rm r}}r_{\rm in}(t).
\end{align}
Here, we have used the first Markov approximation $\gamma_{\rm \ell} = 2\pi\kappa_\ell^2\left(\omega\right)$, $\gamma_{\rm r} = 2\pi\kappa_r^2\left(\omega\right)$ \cite{Gardiner1985}. We also added the internal loss rates $\gamma_{\rm a}$ $\gamma_{\rm b}$ of resonator 2. The two output fields are 
\begin{align}
	l_{\rm out} &= l_{\rm in} + \sqrt{\gamma_{\rm \ell}}a(t), \label{eq:in_out_linear1} \\
	r_{\rm out} &= r_{\rm in} + \sqrt{\gamma_{\rm r}}b(t). \label{eq:in_out_linear2}
\end{align}

Then, we move to the frame rotating with respect to the reference frequency $\omega_{\rm d}$ and define $\Delta_{\rm a}=\omega_{\rm a}-\omega_{\rm d}$, $\Delta_{\rm b}=\omega_{\rm b}-\omega_{\rm d}$, where $\omega_{\rm d}$ is the frequency of the driving field. For steady state solutions, we find
\begin{align}
	S_{11} &= \frac{l_{\rm out}}{l_{\rm in}} = 1-\frac{\gamma_{\rm \ell} \left( i\Delta_{\rm b} + \frac{\gamma_{\rm r}+\gamma_b}{2} \right)}{\left( i\Delta_{\rm a} + \frac{\gamma_{\rm \ell}+\gamma_a}{2}\right)\left( i\Delta_{\rm b} + \frac{\gamma_{\rm r}+\gamma_b}{2} \right) + g^2}, \\
	S_{21} &= \frac{r_{\rm out}}{l_{\rm in}} = \frac{ig\sqrt{\gamma_{\rm \ell}\gamma_{\rm r}}}{\left( i\Delta_{\rm a} + \frac{\gamma_{\rm \ell}+\gamma_a}{2}\right)\left( i\Delta_{\rm b} + \frac{\gamma_{\rm r}+\gamma_b}{2} \right) + g^2}, \\
	S_{12} &= \frac{l_{\rm out}}{r_{\rm in}} = \frac{ig\sqrt{\gamma_{\rm \ell}\gamma_{\rm r}}}{\left( i\Delta_{\rm a} + \frac{\gamma_{\rm \ell}+\gamma_a}{2}\right)\left( i\Delta_{\rm b} + \frac{\gamma_{\rm r}+\gamma_b}{2} \right) + g^2}, \\
	S_{22} &= \frac{r_{\rm out}}{r_{\rm in}} = 1-\frac{\gamma_{\rm r} \left( i\Delta_{\rm a} + \frac{\gamma_{\rm \ell}+\gamma_a}{2} \right)}{\left( i\Delta_{\rm a} + \frac{\gamma_{\rm \ell}+\gamma_a}{2}\right)\left( i\Delta_{\rm b} + \frac{\gamma_{\rm r}+\gamma_b}{2} \right) + g^2}.
\end{align}
Here, we have used the imaginary unit $i$, which is related to the imaginary unit in electrical engineering by $i=-j$. 
We find that all the damping coefficients, $\gamma_{\rm a}$, $\gamma_{\rm b}$, $\gamma_{\rm \ell}$, and $\gamma_{\rm r}$ can be obtained by detuning the two resonators, and measuring the internal and coupling quality factors from the reflection responses $S_{11}$ and $S_{22}$, respectively. 

We find that the scattering parameter $S_{22}$ only depends on the external quality factor $Q_\text{ext,2}  = \omega_\text{b}/\gamma_b$ and the loaded quality factor $Q_{\ell,2}$ of resonator 2, in the limit where resonator 1 is far detuned $\Delta_\text{a} \rightarrow \infty$: 
\begin{align}
S_{22} \approx 1 - \frac{2Q_{\ell,2}/Q_\text{ext,2}}{1 -2iQ_{\ell,2}\Delta_b}.
\end{align} 

On resonance of the second resonator, $\Delta_\text{b} = 0$, we then get
\begin{align}
\label{eq:S22_quality}
S_{22} \approx 1 - \frac{2Q_{\ell,2}}{Q_\text{ext,2}}.
\end{align} 

We extract the loaded quality factor from a fit to the phase $\theta$ of the scattering parameter using
\begin{align}
\theta = \theta_0 + 2 \arctan(2Q_{\ell,2}(1-\omega/\omega_\text{r,2})).
\end{align}

From this we calculate the external quality factor with Eq.~(\ref{eq:S22_quality}).

\begin{table}
  \centering
  \begin{tabular}{c|c}
  parameter & value \\
  \hline
  $\omega_{\rm b}$ &\, $2\pi\times 7.154\,\mathrm{GHz}$  \\ \hline
  $Q_{\rm \ell,2}$ &\, $7.4\times10^3$  \\
  $Q_{\rm int,2}$ & $7.8\times10^3$ \\
  $Q_{\rm ext,2}$ & $1.35\times10^5$ \\ \hline
  $\gamma_{\rm b}$ & $2\pi\times 0.91\,\mathrm{MHz}$  \\
  $\gamma_{\rm r}$ & $2\pi\times 0.05\,\mathrm{MHz}$  \\ \hline
\end{tabular}
\caption{Parameters extracted from the quality factor fitting procedure described in App.~\ref{app:qfactor} near the maximum resonance frequency of resonator 2. We assume the external quality factor to be frequency-independent.}
\label{tab:quality_factor_1}
\end{table}

In our experiment, we detune the two resonators by approximately $ 200\,\mathrm{MHz}$ and measure the reflection response $S_{22}$. To determine the internal and external quality factors from the scattering coefficients, we follow the recipe described in Ref.~\onlinecite{Chen2020} which includes different corrections of the measurement signal. The corresponding fit to the measurement data is shown in Fig.~\ref{fig:InOutFit}. Finally, Tab.~\ref{tab:quality_factor_1} summarizes all determined values. Specifically, we obtain $Q_{\rm ext,2} = 1.35\times10^5$. Comparing this to the internal loss of $7.8\times10^3$, we find that our system is undercoupled. For use in future quantum simulation experiments, beyond this work, an investigation into the causes for the high internal loss will be needed to enable longer polariton life times.

\FloatBarrier
\section{Gain calibration measurements}
\label{app:PNCF}
\begin{figure}
\centering
\includegraphics[width = 0.45\textwidth]{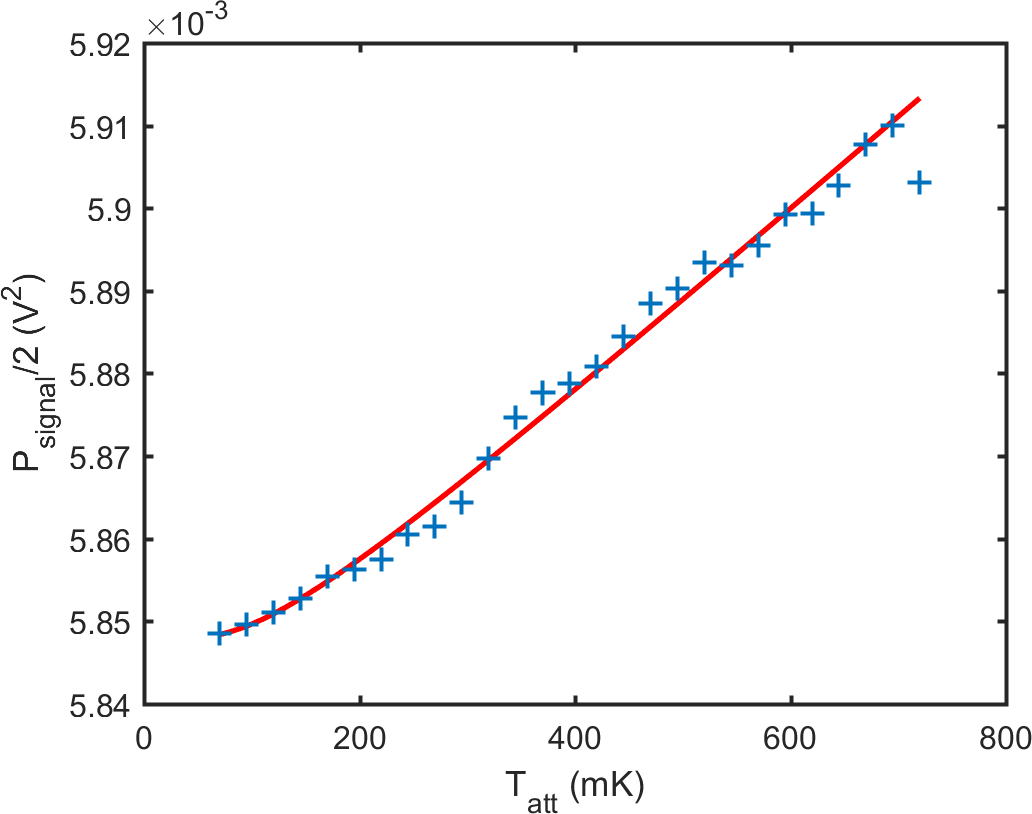}
\caption{Measurement of the signal Power $P_\text{signal}$ in a temperature sweep of the heatable attenuator (blue symbols). The red line is a fit to the data using Eq.~(\ref{eq:PNCF}).}
\label{fig:PNCF}
\end{figure}
As we cannot directly measure the gain of our amplification chain while the cryostat is cold, we perform Planck spectroscopy~\cite{Mariantoni2010,Menzel2012} as a photon number calibration measurement. To this end, we use a heatable attenuator, that emits black body radiation towards the input of the sample. With the resonators far detuned, the signal is reflected to our amplification chain, and we can measure the resulting power with a digitizer card at room temperature. Before digitizing the signal, we convert the signal to an intermediate frequency of \SI{11}{\mega\hertz} with an analog mixer setup. After digitizing, the data is digitally downconverted to DC. As a result, we get the in-phase (I) and quadrature (Q) components of the measured signal as a DC value. We can then write the power of the signal as a function of the temperature of the attenuator
\begin{equation}
\label{eq:PNCF}
P_\text{signal} = \frac{I^2+Q^2}{R} = \frac{\kappa G_\text{cal}}{R}\left[\frac{1}{2}\coth\left(\frac{hf_0}{2k_\text{B}T_\text{att}}\right)+n_\text{noise}\right],
\end{equation}
with the Boltzmann constant $k_\text{B}$, $\kappa = 2R\times BW\times hf_0$, $BW$ the bandwidth of the measurement and $G_\text{cal}$ the total gain of the amplification chain. 
We perform a temperature sweep of the heatable attenuator and measure the resulting output quadrature components, which we can fit to the formula for the black body radiation. The result of this fit is the product $\kappa G_\text{cal}$, which relates the number of photons at the sample to the measured voltages and the photon number $n_\text{noise}$ of the noise of the amplification chain. In our setup, this photon number is dominated by the noise number from the cryogenic amplifier in the chain.
In Fig.~\ref{fig:PNCF}, we show a temperature sweep from \SI{50}{\milli\kelvin} to \SI{800}{\milli\kelvin}. From the fit, we get $\kappa G_\text{cal} = \SI{8.1}{(\milli\volt)\squared\per photon}$ and a noise number $n_\text{noise} = 142$. The high noise number can be explained by aging effects in our HEMT amplifiers. Using the measurement bandwidth of $BW = \SI{2}{\mega\hertz}$, we can calculate the gain of the chain and find $G_\text{cal} = \SI{109.8}{\decibel}$.
As we perform this calibration measurement with a slightly different setup (including an additional downconversion box, containing an intermediate frequency amplifier with significant gain), we cannot directly use this gain measurement for the interpretation in the experiments performed with a VNA. 

In order to estimate the gain of the VNA setup, we measure the gain of the additional room temperature components used in the determination of $G_\text{cal}$. We measure the gain of the downconversion box, $G_\text{box} = \SI{47.3}{\decibel}$, and the gain of an additional rf room temperature amplifier $G_\text{rf-amp} = \SI{24.6}{\decibel}$.
If we subtract these two values from the determined total gain $G_\text{cal}$, we get a value of the gain of the VNA setup $G = \SI[separate-uncertainty=true,multi-part-units=single]{38\pm4}{\decibel}$, which we use in the main text of this paper. The uncertainties stem mainly from the frequency dependency of the gain as the measurement of the gain and of the nonlinearity have been performed at different frequencies (\SI{3}{\decibel}) and the different cables used in the two measurements, which cannot be reliably accounted for in our estimation (\SI{1}{\decibel}). The uncertainty in the gain is the main contribution to the uncertainty of the nonlinearity.
\section{Coupling strength $J$}
In order to extract the coupling strength $J$, we perform a reflection measurement $S_{22}$ or $S_\text{cd}$ at the point of degeneracy at $I_\text{coil} = \SI{19.3}{\micro\ampere}$ and fit a standard input-output formalism to the data. The data has been calibrated by subtracting a background measurement. Visually, $J$ can be seen in this measurement as the splitting between the two resonance dips. From the fit, we extract a coupling strength $J = \SI[separate-uncertainty=true,multi-part-units=single]{7.6\pm0.3}{MHz}$.
\label{sec:J}
\begin{figure}[b]
\centering
\vspace*{0.25in}
\includegraphics[width = 0.45\textwidth]{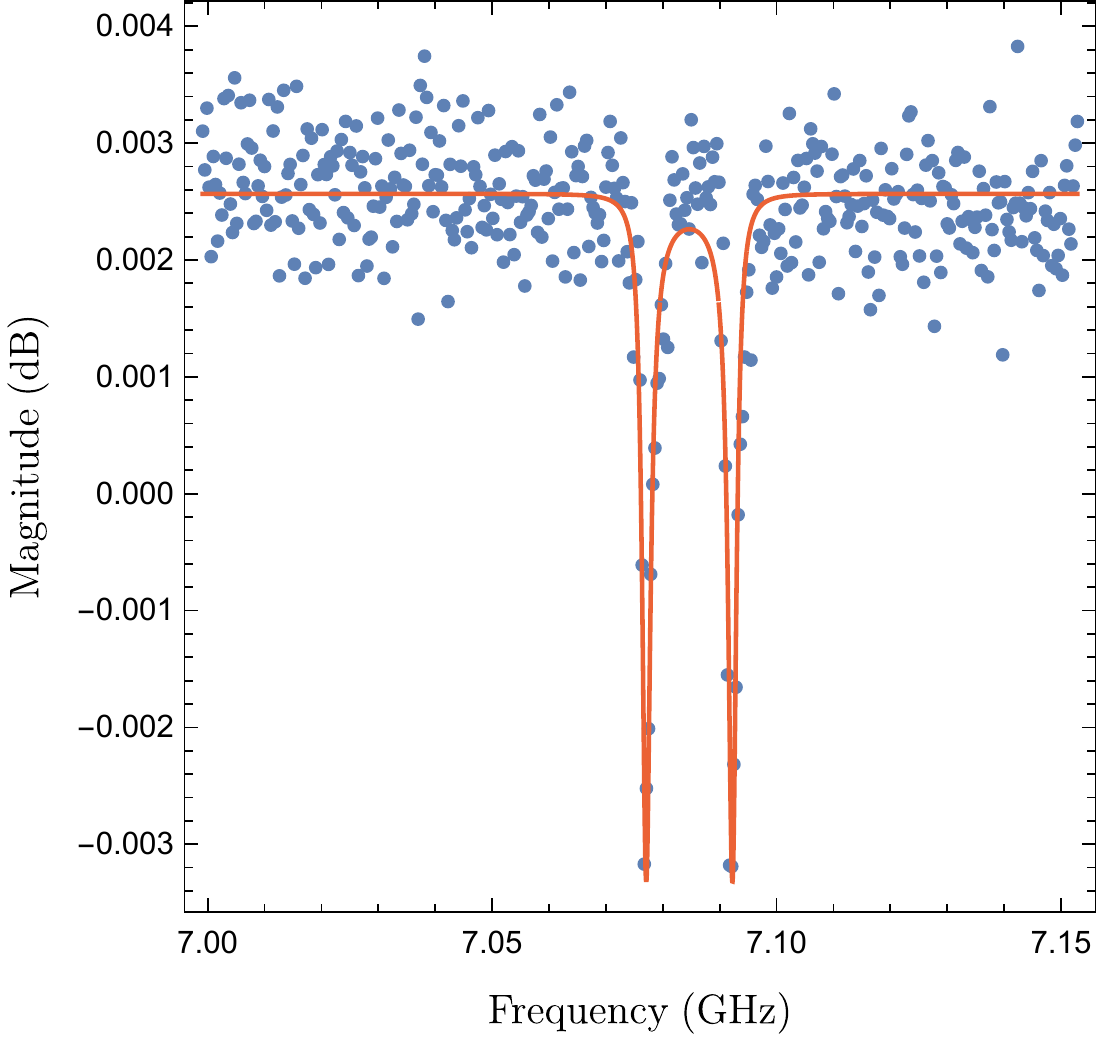}
\caption{Reflection measurement $S_{22}$ (blue dots) at the degeneracy point $I_\text{coil} = \SI{19.3}{\micro\ampere}$. Red line: Fit of a input-output formalism.}
\label{fig:J}
\end{figure}

\end{document}